\documentclass{article}

\usepackage{graphicx}
\usepackage{amsmath}

\newcommand{\onlinecite}{\cite}

\begin{document}

\title{%
  Applications of computational geometry to the molecular simulation
  of interfaces}

\author{
 Florencio Balboa Usabiaga \\
 Daniel Duque  \footnote{ daniel.duque@upm.es
 http://debin.etsin.upm.es/~daniel }\\ 
 Departamento de F\'{\i}sica Te\'orica de la Materia Condensada \\
 Facultad de Ciencias, \\
 Universidad Aut\'onoma de Madrid, \\
 Francisco Tom\'as y Valiente, 7. E-28049 Madrid, Spain.\\[3cm]
}

\begin{abstract}
  The identification of the interfacial molecules in fluid-fluid equilibrium
  is a long-standing problem in the area of simulation. We here propose a new
  point of view, making use of concepts taken from the field of computational
  geometry, where the definition of the ``shape'' of a set of point is a
  well-known problem. In particular, we employ the $\alpha$-shape construction
  which, applied to the positions of the molecules, selects a shape and
  identifies its boundary points, which we will take to define our interfacial
  molecules. A single parameter needs to be fixed (the ``$\alpha$'' of the
  $\alpha$-shape), and several proposals are examined, all leading to very
  similar choices. Results of this methodology are evaluated against previous
  proposals, and seen to be reasonable.
\end{abstract}

\maketitle

\bibliographystyle{unsrt}

\section{Introduction}
\label{sec:intro}

Interfaces, frontiers between homogeneous phases, are ubiquitous in
Nature, playing a fundamental role in the behavior of many complex
systems. Given their inherently more difficult analysis compared to
bulk homogeneous systems, there are still many open questions about
their structure, dynamics, and stability.  The interest in these
structures has been steadily growing in recent years, not only from
fundamental reasons, but also because of their relevance in a great
variety of problem in physical chemistry, biology, and material and
health sciences. Lately, progress in this area has benefited from
advances in computing power and numerical simulation at the molecular
level. Simulation has placed itself in an important pivotal position
between experiments and theory \cite{understanding}. Still, results
from simulations must be careful analyzed, and some features of the
resulting configuration can be hard to obtain, or even define. This is
the case of the interfacial region between two fluids at coexistence,
which is clearly obtained in certain simulations (so-called ``explicit
simulations of interfaces''), but whose precise definition is
elusive.

We present here a novel approach to the analysis of configurations
resulting from a molecular simulation of an interfacial system. We
have divided this introductory Section into two parts; first, we
discuss the physical and mathematical approaches to the interface. We
then introduce some concepts from computational geometry, a branch of
computer science devoted to the study of geometrical problems, such as
the ``shape'' of a set of points. Our proposal is to use methods from
this field in order to analyze the interfacial configurations obtained
by simulation.

\subsection{Theories for the interface}
\label{sec:theories}

In this article we will focus on the most studied and, arguably, the most
important interface, the liquid-vapor interface.  Theoretical approaches to
this problem go back at least to the end of the 19th century, when the seminal
work of Gibbs and van der Waals was presented. The fundamental tenets of these
theories have cast a great influence on this area for about one century. In
particular, a quasi-thermodynamic approach was proposed, in which the
interface is described by a density profile, a function of the normal
coordinate which smoothly goes from the bulk vapor density to the bulk liquid
density \cite{MolThCap}. Mathematically, a position-dependent density profile
$\rho(z)$ is introduced, where $z$ is the coordinate across the interface. A
free energy energy functional of the system is introduced, that assigns a
number to any profile; these theories will be called density functional
theories (DFTs) in this work (even if the usual convention seems to assign
this name to modern ones, excluding van der Waals' original theory). The
equilibrium profile would be the one that minimizes this functional, and the
corresponding value of the functional, the actual value of the free
energy. The surface tension of the interface is given by the excess of this
value over the bulk one (the one with no interface), divided by the area. The
resulting profile is usually monotonic; on the other hand, in the last two
decades, X-ray analysis of liquid surfaces has been employed in order to
characterize their microscopic structure, and results for metals such as
Mercury and Gallium have been interpreted as showing a liquid layering in the
first fluid layers %
\cite{PhysRevLett.74.4444,PhysRevLett.75.2498,PhysRevB.58.R13419,Tostmann1999182}.
In fact, this has also been found by simulation, for models of simple liquids
designed to have low triple point temperatures
\cite{ourlayering,ourlayering2}.
%

A complementary vision of the system at a microscopic level is given
by capillary wave theory (CWT), in which the interface is supposed to
be given by some intrinsic surface (IS) at each moment in time
\cite{MolThCap}. This is a mathematical surface that can be written in
the Monge representation: $z=\xi(x,y)$, and whose precise definition
can vary depending on the author. In any case, CWT concerns itself
with the effect of thermal fluctuations on this IS. One of the
foremost predictions of the theory is the roughness of the source,
which is predicted to (very weakly) diverge with the surface area in
the absence of external, stabilizing fields (e.g., the gravitational
field of the Earth). The main cause of this divergence are the Fourier
modes with long wave-lengths (low wave vectors), which decay too
slowly and thus fail to make the surface roughness
convergent. Mathematically the IS is decomposed in its
Fourier components,
\[
\xi(\mathbf{r})=\sum_\mathbf{q} \xi_q e^{i \mathbf{q} \cdot \mathbf{r}},
\]
where $\mathbf{r}$ is a two-dimensional position vector,
$\mathbf{r}=(x,y)$. The mean square deviation of each of these modes
is given by the theory as:
\begin{equation}
\label{eq:xiq}
\left\langle
\left| \xi_q \right|^2
\right\rangle =
\frac{k_B T}{\gamma_0 A q^2} ,
\end{equation}
where $k_B$ is Boltzmann's constant, $T$ is the temperature, $A$ is
the projected area, and $\gamma_0$ is the bulk surface tension. In the
usual numerical simulations, such as the one presented here, a
stabilizing external field is not present, and surface roughness is
rather controlled by the finite size of the simulation cell, for which
periodic boundary conditions are typically used. In this case the
lowest wave vector is $q=2 \pi /L$ (assuming a square prism of transversal
area $A=L\times L$). The density profile $\rho(z)$ is then in fact
dependent on the area, an effect which is often neglected but has been
reported \cite{sides_1999,ismail_2006}.

It would then be highly desirable to be able to introduce an intrinsic
density profile, 
$\hat{\rho}$, \emph{independent} on the area of the simulation cell
(as long as this is large enough, compared with the correlation
length), which would correspond to the mathematical surface assumed in
CWT. The connection with the usual, average, density profile $\rho(z)$
is given by the ``convolution approximation'': if the intrinsic
profile and the position of the intrinsic surface are taken to be
uncorrelated, then the mean profile is the convolution of the two:
\[
\rho(z)=\int dz' \hat{\rho}(z-z') P(z'),
\]
with the Gaussian probability
\[
P(z)=\frac{1}{\sqrt{2\pi\Delta}} \exp\left(-\frac{z^2}{2\Delta}\right).
\]
The dependence of $\rho(z)$ on the area is clear in the expression
for the width of the Gaussian:
\begin{equation}
\label{eq:cutoff}
\Delta=\sum_q  \left\langle
\left| \xi_q \right|^2
\right\rangle =
\frac{k_B T}{\gamma_0 A} 
 \sum_{\frac{2\pi}{L} < |q|}^{q_u} \frac{1}{q^2} \approx
\frac{k_B T}{4\pi\gamma_0} \log\left(\frac{A}{a}\right), 
\end{equation}
where $q_u$ is a cutoff upper $q$ vector, and $a=(2\pi/q_u)^2$
is the corresponding molecular-sized area.

Recent works have tackled this problem by defining an IS that
separates the two phases
\cite{Chacon_Tarazona_PhysRevLett,Chacon_Tarazona_PhysRevB}. These
methods, which we will call minimum area (MA) methods employ a Fourier
description of the surface (like in the original CWT). The Fourier
components are determined by a requirement that the surface passes
through a set of surface molecules, termed ``pivots'', while having a
minimum area. From an initial set of pivots a surface is thus
constructed; the molecule that is closest to the surface is
incorporated as a new pivot, a new surface is constructed, and the
process is iterated until a target number of pivots is reached.  With
this procedure, the interfacial region has been seen to have a much
richer structure, with intrinsic density profiles that are similar to
the radial correlation function of the liquid and show clear layering
even at temperatures close to the critical point.  The new definition,
besides shedding a new light on the microscopic structure, allows the
study of dynamical features, such as diffusion, which is relevant e.g.
for surface reaction dynamics \cite{Duque_Tarazona_Chacon_2008}.

\subsection{Computational geometry}
\label{sec:geometry}

The analysis of disordered media has grown in importance lately, both
within physics and in many other areas; we will be using one of the
most well-known techniques, Voronoi tessellations and Delaunay
triangulations, together with the lesser-known
$\alpha$-shapes \cite{edelsbrunner,EM_1994,okabe}. Given $N$ points in
three-dimensional space, the Voronoi polyhedron associated with each
point is the region in space that is closer to the point than to any
other point. It is also the smallest polyhedron formed by the
bisecting planes of the lines joining the point with all the
others. This concept and procedure are well known in solid state
physics, where the object is called the Wigner-Seitz cell, the
primitive cell of a crystal structure, but the Voronoi cells can be
defined for disordered media.

From the Voronoi diagram, or tessellation, which is the set of all
polyhedra, one may obtain the Delaunay triangulation, by joining the
points that share a common facet. (One may call this a
tetrahedralization in three-dimensions, but triangulation is the
preferred name; indeed, the cells of the space partition are
tetrahedra, but their facets are triangles.)  This triangulation
provides a precise definition of ``neighbor'': points directly
connected through the triangulation are first neighbors, point needing
two connections are second neighbors, and so on. Even if this
description of Delaunay triangulation builds from the Voronoi diagram,
it is actually more convenient computationally to do the opposite, and
directly compute the triangulation. Besides, the Delaunay
triangulation itself satisfies a number of interesting mathematical
properties, the chief one being that it is unique for ``almost'' any
set of points, and that the tetrahedra in it satisfy the \emph{empty
  sphere condition}.  That is: for any tetrahedron, defined by four
points which are mutually neighbors in the triangulation, the sphere
that passes through them all contains no other point in its
interior. This criterion is, in fact, a key part in computing these
triangulations. Previous work in this area that has made use of
Voronoi diagrams includes examination of glassy and disordered
systems \cite{Medvedev_1990,Gil_Abascal_1993}, neighbor
statistics \cite{kumar_2005}, and configurational
entropy \cite{kumar_2005_2}; there are some works devoted to the study
of the liquid-vapor interface by Voronoi
tessellations \cite{fern_2007,fern_2007_2}, but these focus on bulk
quantities, such as densities, not on structure.

Another construction, which is the main novelty of this work, is the
$\alpha$-shape \cite{edelsbrunner,EM_1994,okabe}. This shape has been
historically introduced precisely in order to define the ``shape'' of
a given set of points, and it also provides a definition of the
``border'' points of the shape.  These concepts have been applied to
three dimensional scanning data, and also to biochemistry \cite{duke},
and seem ideally suited to our particular problem of finding the
``outside'' of a liquid phase (at least, when the gas is rarefied
enough). For a given set of $N$ points, there exist many
$\alpha$-shapes, which may be obtained by the value of the parameter
$\alpha$. Let us define a distance $R$ that fixes the value of
$\alpha$; depending on the author the relation between the two
varies. We will take the choice of the (\textsc{Cgal})
project \cite{cgal} : $\alpha=R^2$, but for others 
$\alpha=1/R$ and our shapes correspond to negative values of $\alpha$.

\begin{figure}
\centering
\includegraphics[width=0.8\textwidth]{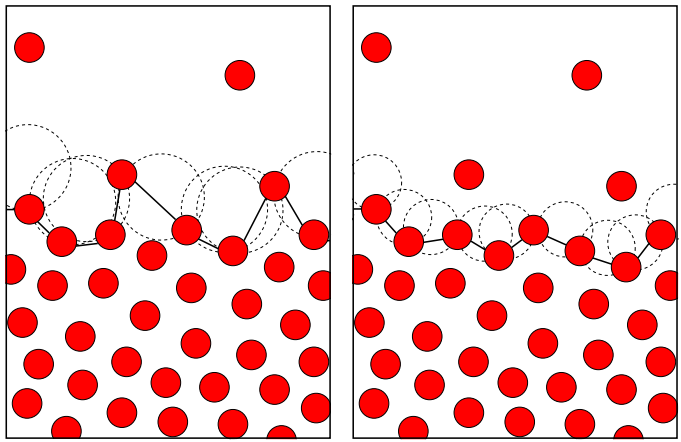}
\caption{%
\label{fig:schematic} Schematic diagram of the
  $\alpha$-shape construction.  For the sake of simplicity, a two-dimensional
  construction is shown. Left: a set of particles (red circles) is endowed
  with an $\alpha$-shape (whose border is plotted with a solid line)
  corresponding to ``scoops'' of radius $R$; dashed circles represent scoops
  with maximum penetration (touching two particles). Right: for the same
  particles, a smaller scoop radius produces a different $\alpha$-shape. }
\end{figure}

An intuitive definition of the procedure is as follows \cite{EM_1994}.%
One may think of the whole of space as filled out with ice cream, with
$N$ chips in it.  A spherical ``scoop'' of radius $R$ carves out balls
of ice cream, without removing the chips. (Notice we may also carve in
between the points, which breaks the analogy with a real scooping
process, where the scoop must arrive ``from the outside''.)
The result of the process will be a region of space which may have a
complicated shape, which is by bounded caps, arcs and points. If we
straighten all caps to triangles and all arcs to line segments, we end
up with an $\alpha$-shape. This object has a well-defined border: the
set of bounding points --- these are the points that have been
``reached'' by the scoop but are still part of the $\alpha$-shape
(i.e. the spherical scoop has been able to remove some of the ice
cream around them, but not \emph{all} of it).  This set of points
define the $\alpha$-shape border. Notice the procedure also introduces
a definition of neighborhood: three points are neighbors if a scoop
has reached the three of them. (This is a description of three
dimensional $\alpha$-shape, their two-dimensional analog also exists
and is easier to visualize, but has a lesser applicability.  Also,
note that some authors choose to define an $\alpha$-shape by what we
here call its border, which may lead to confusion.)  In Figure
~\ref{fig:schematic} we show a schematic diagram of the construction;
for a set of points we show how the choice of $R$ affects the final
$\alpha$-shape. For the sake of simplicity, a two-dimensional
construction is employed.

Two limits will perhaps clarify the procedure. If $\alpha$ is very
small, all of the points will be reached; the resulting shape is,
therefore, all of the $N$ points --- but none of them will belong to
the ``border''. The other limit is perhaps more interesting: for
values of $\alpha$ very large, the scoop takes half-spaces out, and
can only reach the points that protrude from the surface. In this
limit, the resulting $\alpha$-shape has as its border the very
well-known \emph{convex hull} of a set of points.  We have made use of
the Computational Geometry Algorithms Library
(\textsc{Cgal}) \cite{cgal}, which provides highly efficient codes for
these kind of structures. The $\alpha$-shape is evaluated in this
library at little additional computational cost once a Delaunay
triangulation is computed: the empty sphere condition permits to
associate a given radius to each tetrahedron, below which the scoop
will be able to ``get in''. It is therefore a matter of bookkeeping to
find the accessible points given some $\alpha$ value. More
technically, the $\alpha$-shape is build in the (default) regularized
mode, which eliminates isolated faces, and the border points are
identified as the ``regular'' points \cite{cgal:dy-as3-07}.

This procedure is related to a recent proposal by L.B. P\`artay and
collaborators \cite{partay_2008}, in which they employ spheres
approaching the surface in the normal direction, from the vapor
side. This method and the one examined here are likely to provide
similar results at low temperatures. At higher ones, on the other
hand, their method will be hampered by molecules that are either in
the vapor phase or loosely attached to the surface.  Moreover, it will
miss overhangs, whose role may be important at temperatures close to
the critical point (note that this also applies to the MA methods,
since they assume a surface defined mathematically by a single
function $z=\xi(x,y)$, which cannot present overhangs.) In this sense,
the current method can be considered more robust and reliable;
moreover, it can be computationally more efficient than either the
methods of Ref. \onlinecite{partay_2008} or MA methods (we will
discuss efficiency in the next section).  In addition, this method is
not restricted to flat interfaces and can readily be applied to other
interfaces, such as curved ones. This may be very useful in studies of
nucleation (or cavitation), where a knowledge of the interfacial
properties of droplets (or bubbles) may be desirable --- this method
may also be employed for micelles and other supramolecular
aggregates. In fact, for most results we do not need to distinguish
between an ``upper'' and ``lower'' interface, since the method selects
both of them automatically. It is also easy to label each of the
particles in the system as being part of the IS, next to it, second
next to it, \textit{etc}, since the Delaunay tessellation provides a
well defined definition of neighborhood (as explored in the context of
correlation functions in Ref. \onlinecite{kumar_2005}).

Another relevant reference is the work by J. Chowdhary and
B.M. Ladanyi \cite{chowdhary_2006}. In their work, the selection of
surface molecules is alleviated because they consider binary
interfaces between water and hydrocarbons. Since the two components
are nearly immiscible, a simple definition of proximity to the
opposing species can be employed to identify pivots. They go forward
and construct an IS, a step not taken in P\`artay
\textit{et al}  \cite{partay_2008}. In a spirit similar to this
article, concepts of computational geometry are employed: the pivots
are projected onto the plane of the interface, and a Voronoi
tessellation is constructed (this procedure is closely related to the
one termed ``Delaunay Terrain'' in the \textsc{Cgal}) documentation
\cite{cgal}.) This allows an association of each non-surface molecule
with a surface one, from which intrinsic profiles may be computed. In
line with the MA results, structured intrinsic profiles are found for
water, even when average profiles are monotonous (for hydrocarbons
both are structured, the intrinsic ones more so.)

The method as implemented in \textsc{Cgal} is suitable only for
``open'' boundary conditions. E.g., an additional ``point at
infinity'' is introduced when computing triangulations. Our system
features periodic boundary conditions, which are not yet implemented
in \textsc{Cgal} (a project to include them is underway
\cite{monique}). We therefore choose the clumsy, but effective,
procedure of replicating slabs of our cell, then neglecting the points
which are outside the original cell. We have made sure these slabs are
wide enough, $4\sigma$, for the boundary conditions to be well
represented.

\section{Methodology}
\label{sec:methodology}

We simulate a fluid with $2592$ particles interacting through a
Lennard-Jones (LJ) potential:
\begin{equation}
\label{eq:LJ}
u(r)=
\epsilon
\left\{
  \left(
  \frac{\sigma}{r}
  \right)^{12}
-
  \left(
     \frac{\sigma}{r}
  \right)^6
-
  \left(
     \frac{\sigma}{r_\mathrm{c}}
  \right)^{12}
+
  \left(
     \frac{\sigma}{r_\mathrm{c}}
  \right)^6
\right\}.
\end{equation}
Interactions are truncated at a cutoff radius of
$r_\mathrm{c}=3.02\sigma$.  Particles are confined in a rectangular
cell of dimensions $L_x=L_y=L=10.46\sigma$ and $L_z=90\sigma$; periodic
boundary conditions are applied in all directions. Standard molecular
dynamics simulations are carried out, with a time-step $dt = 4.56
\times 10^{-3} \sigma\sqrt{m/\epsilon}$ (in reduced units),
using the software package \textsc{dl\_poly} \cite{dlpoly}. After an
equilibration period of $10^6$ steps, particles form a liquid slab in
the $x$-$y$ plane, surrounded by vapor. The temperature is set at
$k_BT = 0.678\epsilon$ by a Nose-Hoover thermostat. This has been
found to be the triple point temperature for this truncated LJ
potential, and hence the lowest temperature at which the liquid is
thermodynamically stable. A production run of $50000$ steps is carried
out, in the microcanonical ensemble, in order to avoid possible
artifacts of the thermostat on the dynamics (which will be important
for one of the methods explained below). Interfacial dynamics is slow
enough to let us analyze $1$ configuration out of $10$, so we have to
sample $5000$ configurations.

This choice of parameters (specially, number of particles and cutoff radius)
leads to a very fast computation, that may be carried out even in a standard
modern laptop computer. The reason for carrying out this simple simulation is
that we will compare our results with MA. The later technique is quite heavy
computationally, so that its configurational analysis ends up being more
time-consuming than the simulation itself. For this number of particles, the
MA analysis takes a CPU time about $10$ times longer than the simulation. The
current procedure takes a time comparable to the simulation, depending on the
information requested (about the same for the number of pivots, twice as much
to build the intrinsic profile, $10$ times as much for the Fourier analysis
described below.)  Moreover, the MA procedure grows with the number of
particles as $N^2$ (if the surface area is scaled suitably, so that the area
grows as a power $2/3$ of the number of particles), whereas the current
procedure grows as bad only in the worst case: typically, it grows only as
$N\log(N)$ [the Fourier analysis scales as $N^{2/3}\log(N)$ .] We have carried
out additional runs in order to get error bar estimations for some
calculations, as will be indicated.

In order to define the border between the liquid and vapor phases,
$\alpha$-shapes are employed. The particles that define this border,
which will be the outmost liquid layer, will be called ``pivots'',
keeping the standard nomenclature.  The idea is to obtain these pivots
as the points belonging to an $\alpha$-shape. Notice that the
resulting shape is not a smooth surface, as in previous approaches,
but a triangulated one.

\begin{figure}
\centering
\includegraphics[width=0.8\textwidth]{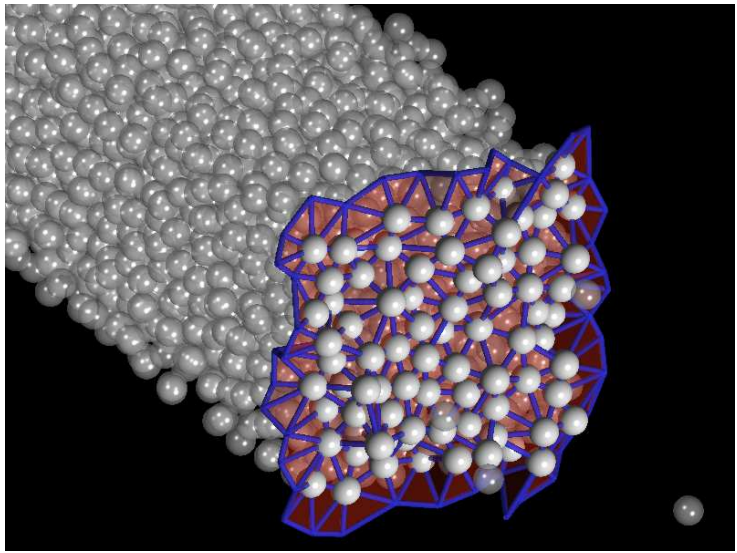}
\caption{%
 \label{fig:snap} (Color online) Snapshot of one of our configurations.
  Molecules are pictured as white balls, translucent unless they belong to the
  intrinsic surface, in which case they are solid. The $\alpha$-shape border,
  for $\alpha=R^2$, $R=1.2$, is the red triangulation, with edges in blue.
}
\end{figure}

A first issue arises from the fact that there are isolated particles
that belong to the gas that will be included as part of this shape for
any reasonable value of $\alpha$. This is easily taken care of by
choosing the points that have surface neighbors in the sense given
above (having a sphere of radius $R$ touching all three), and marking
the rest as isolated. In the \textsc{Cgal} implementation, this is
taken care of by choosing the regular points in the regularized mode,
as commented above. This procedure, indeed, is similar to the
percolation analysis that is employed as a first step in MA
approaches. In Fig.~\ref{fig:snap} we show a snapshot of our
configuration and the resulting $\alpha$-shape, for a typical choice
of $\alpha$ explained below.

\section{Results}
\label{sec:results}

Of course, the main task is to determine the optimal value of the
scoop radius, $R$, or equivalently the right $\alpha$-shape (since
$\alpha=R^2$.)

At variance with many previous applications, the system under
consideration has some well defined length scales. For dense phases,
such as the liquid, the interparticle spacing is always close to the
distance at which the potential has its minimum ($2^{1/6} \sigma$ for
the LJ fluid). Some of the particles may be slightly further, but they
can never be much closer, since at some distance hard-core repulsion
sets in (around $\sigma$ for the LJ fluid). This means that a choice
of $R<\sigma$ will result in the selection of all of the particles as
belonging to the $\alpha$-shape, a nonsensical result from a physical
point of view. Values of $R$ too high, on the other hand, result in
few pivots being selected, and a nearly flat shape (with our boundary
conditions, one outmost molecule at each interface will be selected as
a pivot, with horizontal interfaces.)  In practice, we will need to
consider a range $\sigma < R < 2\sigma$, or
$\sigma^2 < \alpha < 4 \sigma ^2 $.

There are several ways to determine this value, which
should yield the same result (within error intervals):
\begin{itemize}
\item examination of the profiles,
\item pivot dynamics, and
\item comparison against previous, reliable, results.
\end{itemize}
These three paths are described in the next subsections.

\subsection{Structure}
\label{sec:structure}

\begin{figure}
\centering
\includegraphics[width=0.8\textwidth]{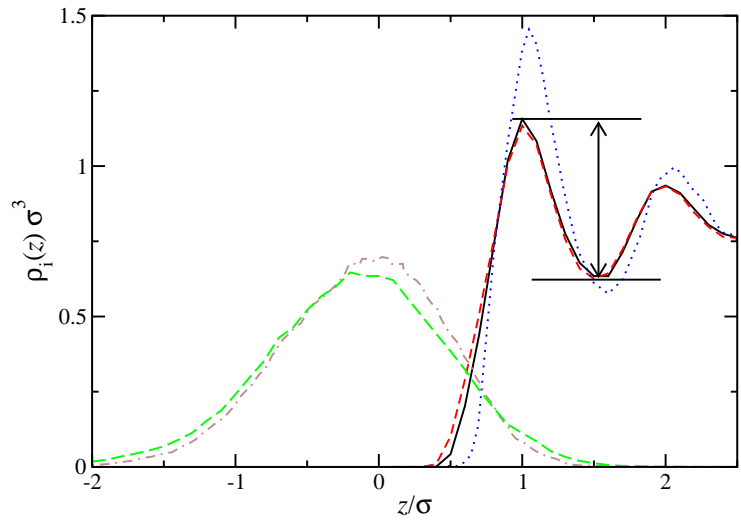}
\caption{%
 \label{fig:intrinsic}
  Density profiles as functions of the coordinate across the
  interface.
  Solid black line: intrinsic density profile $\rho_i(z)$ obtained by
  the current method, for $R=1.20\sigma$; dashed red line:
  $\rho_i(z)$ obtained by the current method, for $R=1.30\sigma$; blue
  dotted line: MA $\rho_i(z)$ taken from Ref.~\onlinecite{IS}. The
  vertical arrow marks the difference that serves as a measure of
  order.
  Green long-dashed line: pivot profile $\rho_s(z)$ obtained by the
  current method, for $R=1.30\sigma$; brown dot-dashed line: MA
  $\rho_s(z)$ taken from Ref \onlinecite{Duque_Tarazona_Chacon_2008}.
  The two later curves are normalized to unity.
}
\end{figure}

The first route is through careful examination of details of the
profiles. Several profiles can be obtained: the average profile,
$\rho(z)$, the pivot profile $\rho_s(z)$, and the intrinsic profile
$\rho_i(z)$. The first is of course independent of the procedure to
obtain the IS. The overall shape and decay of the other two are
features that have been used in order to determine the best value of
the parameters. As discussed in Refs. \onlinecite{IS, our_new}, this
is quite painstaking and fine-tuning is difficult.  For example, in
Fig. \ref{fig:intrinsic} we plot results for $R=1.20\sigma$ and
$R=1.30\sigma$ which, as we will see, are close to optimal; it is
clearly difficult to judge which profile is ``best.'' We therefore
focus on a specific detail of the profiles: the difference between the
height of the first peak and the next trough (the one between the
first and second peaks). This distance will be a measure of the
structure of the intrinsic profile.

In principle, a more structured profile should be preferable. The
argument behind this claim is partly circular: given that we assume
the existence of some IS which provides some intrinsic
density profile (much rougher than the average one), the best
criterion is the one that results in the rougher profiles (always, of
course, within some logical physical limits) \cite{pedro}. In fact,
this criterion enables us to compare different methods, as we will see
in \S\ref{sec:previous}. Of course, this point of view should not be
pushed too far, and profiles that oscillate unphysically, or present
other unlikely features, should be discarded. Ultimately, a comparison
with results obtained by other methods (such as experiments) would
favor one method or another, but for the time being this comparison is
not possible.

\begin{figure}
\centering
\includegraphics[width=0.8\textwidth]{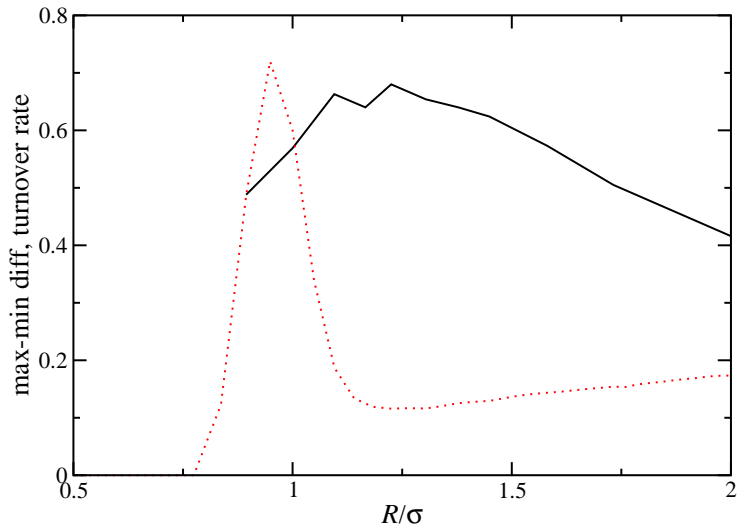}
\caption{%
 \label{fig:struct} (Color online) Solid black line: difference
  between heights of the first peak and next trough of the intrinsic
  profile (see arrow in Fig. \ref{fig:intrinsic} for a particular
  case), versus scoop radius $R$. Dotted red line: Pivot turnover rate
  (arbitrary units) versus scoop radius $R$.
}
\end{figure}

In Fig. \ref{fig:intrinsic} we draw vertical segments that represents
the measure of structure. for two particular choices; the collected
values are plotted in Figure~\ref{fig:struct}. Apart from some noise,
an optimal value of $R$ of about $1.2\sigma$ is clearly identified ---
this could be anticipated, as it is close to the mean interparticle
distance in the bulk liquid, which in turn is similar to the value at
which the LJ potential has its minimum.

\begin{figure}
\centering
\includegraphics[width=0.8\textwidth]{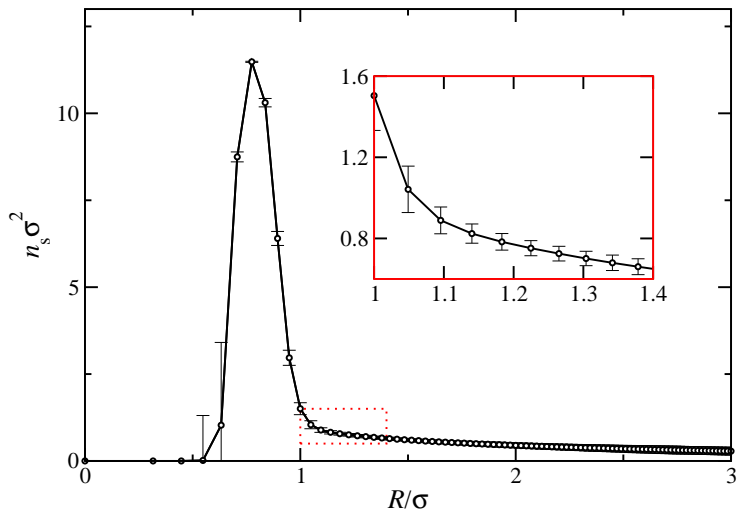}
\caption{%
\label{fig:n_s} (Color online) Surface density (pivots per unit
  area) versus scoop radius $R$. The error bars correspond to the
  variance of the density due to different samples. Inset: blow-up of
  relevant region, shown as a red dotted box in the main graph.
}
\end{figure}

In Fig. \ref{fig:n_s} we show the calculated surface density $n_s$, an
important surface characteristic which is simply defined by the number
of pivots per unit projected area (i.e., the nominal area, not the
area of the IS). The inset shows a blow-up of the area
most relevant for this study. For $R=1.2\sigma$ we can read off the
value $n_s=(0.77 \pm 0.2)\sigma^{-2}$, to be compared with previous
estimate \cite{our_new} of $0.80\sigma^{-2}$.

\subsection{Dynamics}

As shown recently \cite{our_new}, dynamics can provide a reliable,
independent way of obtaining the optimal parameters for the IS, apart
from the obvious interest in dynamical features of the interface
\cite{Duque_Tarazona_Chacon_2008}. A key quantity in this regard is
the \emph{turnover rate}: the number of molecules per unit area and
per unit time that enter the IS (i.e., that become pivots). This
number also equals the number of molecules that leave the IS, on
average (or, exactly if the number is fixed, as in most recent MA
approaches \cite{IS,ISM_aplications,ISM_aplications2}.) The pivot
residence time is physically influenced by the fluid dynamics at the
molecular level, but it is also tied to the specific choice of IS. If
the number of pivots is too high, there will be a high number of them
that enter and leave the IS at each time-step: many of these pivots
should not be included in a physical IS, and this high turnover is
spurious. If, on the other hand, the number of pivots is too low, the
IS will change its shape a great deal, jumping from a set of pivots to
another one, thus causing a high turnover. The best parameter is, in
principle, the one for which a minimum in turnover rate is found. This
argument applies most directly to procedures in which the number of
pivots is fixed, but it may be applied to other parameters, such as
$R$, that indirectly affect this number (with a clear dependency, as
shown in Fig. \ref{fig:n_s}). In Fig. \ref{fig:struct} we plot this
turnover rate, which identifies a value of $R=1.23\sigma$ --- this is
in rather good agreement with the previous value of $1.2\sigma$.  The
corresponding surface density is $n_s=(0.74 \pm 0.2)\sigma^{-2}$.

A related criterion, not strictly dynamical in nature, is to check the
variance implied in the error bars of Fig. \ref{fig:n_s}. This
quantity shows a minimum at $R=1.25\sigma$, in very good agreement
with the previous value (there is a deeper minimum at $R\approx
0.75\sigma$, but this range is unphysical).

\subsection{Comparison to known results}
\label{sec:previous}

We may compare the resulting intrinsic density profiles with the ones
from the MA method. In Fig.~\ref{fig:intrinsic} we also show
(bell-shaped curves) the density profile of surface molecules
(pivots), $\rho_s(z)$, compared with a previous MA profile. The
parameter $R$ is set to $1.3\sigma$, which, as has been explained, is
close to optimal. Our current results show a slight decrease of
central pivots and an increase of peripheral ones. This means the
$\alpha$-shape is including molecules slightly toward the two bulk
phases, which are not part of the MA list, and neglecting some
interfacial ones. This does not seem desirable, but the two curves are
seen to be quite close in shape.

On the other hand, a comparison of our intrinsic density profile
$\rho_i(z)$, already discussed, with the MA one (both included in the
same Figure, with a solid and dotted lines) clearly shows that the
current method, despite its elegance, still performs worse than the MA
one. Keeping the criterion employed in \S\ref{sec:structure}, a
stronger structure signals a better method, and our intrinsic
profiles, while showing distinct layering, are smoother than the MA
ones. We will discuss possible improvements in the Conclusions
section, but for now let us take this as an indication that MA results
may be taken as close to the ``optimal'' ones.

\begin{figure}
\centering
\includegraphics[width=0.8\textwidth]{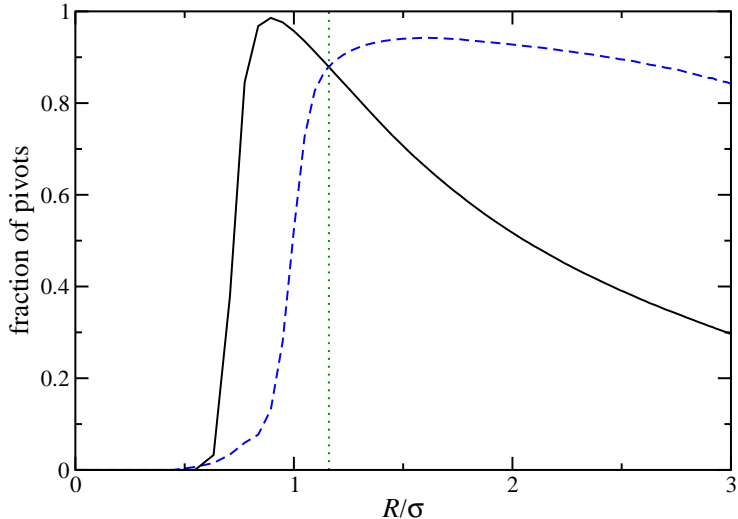}
\caption{%
\label{fig:pivot_comp} (Color online) Fraction of pivots versus
  scoop radius $R$.  Solid black line: fraction of MA pivots that are
  part of our pivots. Dashed blue line: fraction of our pivots that
  are part of MA pivots. The dotted green vertical line provides the
  intersection of the curves, at $R=1.16\sigma$.
}
\end{figure}

We may now use this idea in order to rapidly establish the optimal
values of $R$, by comparing against the MA results. In
Fig.~\ref{fig:pivot_comp} we plot (solid line) the fraction of MA
pivots that are part of our
pivots.
The agreement is fairly low for high values of $R$, since very few
pivots are selected in this case (recall only the most exposed
molecules are selected for very high values of $R$). The agreement
then increases, and goes to almost $1$ at values around $\sigma$. This
means all MA pivots are included in our set, but this is hardly
surprising, since around this range most molecules are selected as
borders of the $\alpha$-shape, including of course most MA pivots (see
the high peak in Fig. \ref{fig:n_s}). The information can therefore be
completed by considering the fraction of our pivots that are contained
in the MA list. This curve is also plotted in
Fig.~\ref{fig:pivot_comp} with a dashed line.  The behavior is
qualitatively similar, but this time the maximum is much more rounded,
and occurs at higher values, around $1.5\sigma$.  Since the maxima do
not coincide, we may take the value at the intersection of the two
curves as the best compromise. At this value, about $88$\% of the MA
pivots are included in our list, and $88$\% of our pivots are
included in the MA list. This value turns out to be $R=1.16\sigma$,
again in good agreement with previous estimates. Alternatively, this
crossing is the point for which both methods coincide in the value of
the surface density, $n_s=0.80\sigma^{-2}$. (If a set A contains a
fraction of elements common with another set B, and B contains the
same fraction of elements common with A, both sets must have the same
size.) An alternative procedure would therefore be to select the value
of $R$ for which a previously known surface density is recovered.

\begin{figure}
\centering
\includegraphics[width=0.8\textwidth]{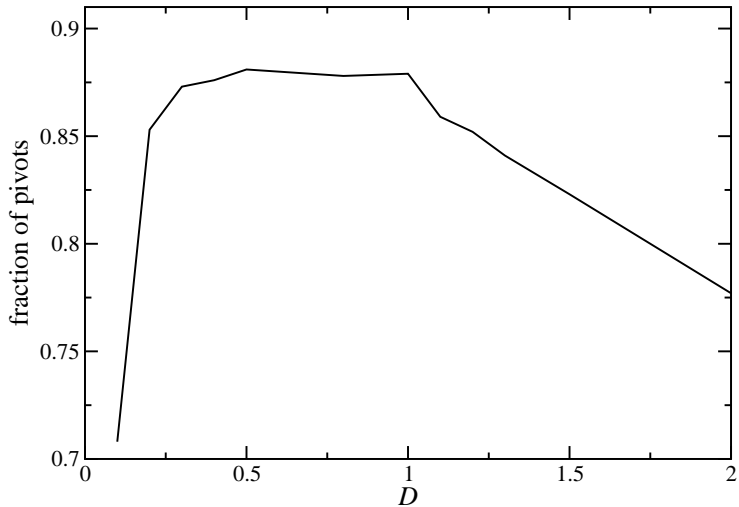}
\caption{%
 \label{fig:pivot_comp2} Fraction of MA pivots that are part of
  our pivots versus anisotropy parameter $D$.
}
\end{figure}

Given the strong anisotropy of the interface in the direction across
it, we have also explored the possibility that one may obtain better
results with the use of ellipsoidal scoops, instead of spherical
ones. (The same numerical libraries may be used, since what we
effectively do is to compress the system in the $z$ direction instead,
while keeping the scoops spherical.)  The scoops would be defined by
\[
(x/R)^2 + (y/R)^2 + (z/D R)^2 =1,
\]
with two parameters, $R$ as before, and $D$ measuring the anisotropy.
This new parameter, $D$, is explored in
Fig.~\ref{fig:pivot_comp2}. For each value of $D$ the optimum value of
$R$ is found, and the curves therefore implies different values of the
later, which is $1.16$ at $D=1$. It is apparent that the best value
turns out to be very close to $D=1$.
This is also the case for the methods above (result not shown): the
difference in height is maximal for this value, the minimum of the
turnover rate is shallower for other values of $D$, and does not
coincide with the minimum of the variance.  Hence, the idea of using
ellipsoidal scoops can be rejected, at least for this simple liquid.

Since our triangulated mesh is a perfectly defined mathematical
entity, a full Fourier transform of the surface can be carried out, at
an arbitrary level of detail. This is at variance with the MA
procedure, in which a higher cutoff wave-vector $q_u$ must be
introduced in order the method be mathematically tractable. While
there is a physical reason for this wave-vector being close
$q_u=2\pi/\sigma$ (as usually done in CWT, see Eq. \ref{eq:cutoff}),
finding its precise value has been delicate.

\begin{figure}
\centering
\includegraphics[width=0.8\textwidth]{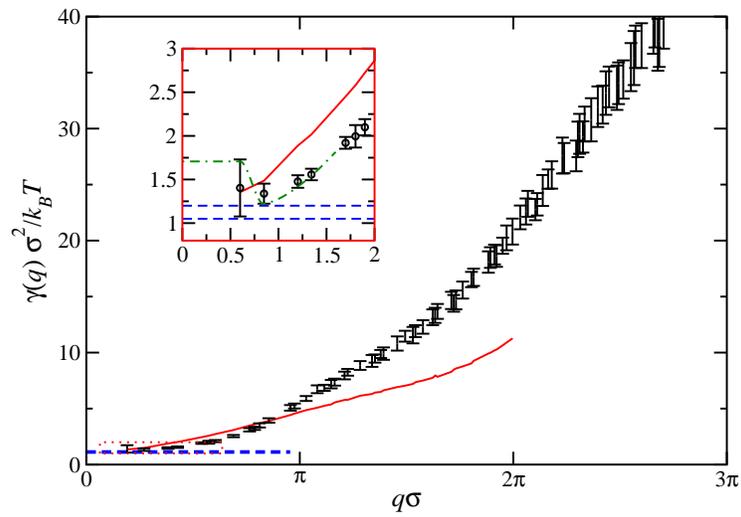}
\caption{%
 \label{fig:surftens} (Color online) Wavenumber-dependent surface
  tension $\gamma(q)$ in reduced units.  Black bars: current results,
  with error bars; solid red line: MA results \cite{datos_chacon} for
  $n_s=0.8\sigma^{-2}$, $q_u=2\pi/\sigma$; dashed blue line: range of
  values for the bulk surface tension, obtained by the method of
  \cite{Duque_Pamies_Vega_2004}. The inset shows a blow-up of the
  low-$q$ region, shown as a red dotted box in the main graph. In it,
  a possible deepest-minimum situation is plotted with a green
  dot-dashed line.
}
\end{figure}

In Fig. \ref{fig:surftens} we compare results from the Fourier
analysis of our IS surface with recent MA results. In particular, we
compute the very important surface tension spectrum; CWT establishes
the mean square fluctuations of Fourier mode given in Eq. \ref{eq:xiq}
above. This expression is seen to be only valid at low $q$ values; at
higher values, the $q$ dependence is taken care of by defining a
$q$-dependent surface tension:
\[
\gamma(q) \equiv
\frac{k_B T}
 {
   A q^2
   \left\langle
     \left| \xi_q \right|^2
   \right\rangle
 }.
\]

The main argument behind CWT hinges on this function increasing with
$q$, otherwise the surface would be unstable against high-$q$
perturbations.
In order to refine our precision, we have carried out $4$ additional
runs, and used results from the two interfaces present in the system,
in order to obtain $8$ values for each value of $q$. We have used
their mean value as our measured quantity, and their standard
deviation as its error bar. The error bars are seen to be increasingly
smaller at lower values of the wave-vector, with statistical errors
increasing at higher values

In the same graph we include previous MA results for this system
\cite{datos_chacon}, and the bulk surface tension. For the later
quantity, a new simulation has been carried out at this particular
temperature and potential cutoff; we indicate the range of the surface
tension, that is, a band centered in the mean value and with a width
given by the error, calculated as in
Ref. \onlinecite{Duque_Pamies_Vega_2004}. Our curve is qualitatively
similar to the MA one: both tend to the bulk value in the limit
$q\rightarrow 0$ and then increase, and both increase further for high
values of the wave-vector, in accordance with the CWT
thesis. Quantitatively, they differ in the values of $\gamma(q)$, with
the MA curve starting out with a higher slope, but then growing slower
than our curve at higher values of $q$.

At this point, we must mention that our results share a feature with
those from MA: the absence of an intermediate minimum between the
limits of very small and very large wave-vectors. To be more precise,
our error bars would in principle allow for a minimum around
$q=1/\sigma$, as shown in the Figure's inset, where a possible
deepest-minimum situation is depicted. However, the inset also shows
that this minimum is not likely to be present at all, given the
range of the values for the bulk surface tension.


The presence of a minimum would imply the enhanced amplitude of
capillary waves with wave-vectors around it. This possibility has been
predicted theoretically by Mecke and Dietrich \cite{Mecke_Dietrich},
improving previous efforts \cite{PhysRevE.47.1836}. The theory was
later expanded \cite{lovett:10711, Hiester_Dietrich_Mecke, Stecki,
  segovia-lopez:021601} and received support from simulations
\cite{vink:134905}. Remarkably, experimental work has also confirmed
this prediction
\cite{nature,PhysRevLett.90.216101,PhysRevLett.92.136102,Lin2005106,PhysRevB.72.235426}.
However, other experimental teams have pointed out technical
difficulties in the interpretation of the experimental
data \cite{Pershan2000149,PhysRevB.69.245423}. Likewise, the main
theoretical framework has been recently challenged
\cite{tarazona_checa_chacon}. There are indications (see recent
Ref. \onlinecite{blokhuis:014706}) that this disagreement stems from
whether the IS is defined from the molecular positions (as in this
work), or from molecular distributions (such as a set of Gibbs
dividing surfaces).

\section{Conclusions}
\label{sec:conclusions}

We have presented an application of the $\alpha$-shape concept, taken
from the field of computational geometry, to the identification of the
intrinsic surface of a liquid-vapor interface. Compared with the
minimum area method, the method may have a lower performance, as
indicated by the less structured intrinsic density profiles. However,
the method is elegant and computationally very simple. Both from a
computational point of view and, more importantly, from a mathematical
one. The resulting IS is a triangulated mesh, which is a perfectly
defined mathematical entity. As a result, a full Fourier transform of
the surface can be carried out, at an arbitrary level of detail.
Our results for the surface tension spectrum are qualitatively similar
to the ones from MA, even in quantitatively different. In particular,
no sizable minimum in the surface tension is observed.

We therefore feel that the current method shows a number of
interesting features. Even if still inferior to the MA method, further
developments could refine it and make it comparable. It is
interesting, in this regard, to incorporate some of the basic ideas of
MA in the procedure: either minimum-area requirements, or incremental
adding of pivots to an existing surface.

\section{Acknowledgments}

Financial support for this work has been provided by the Direcci\'on
General de Investigaci\'on, Ministerio de Ciencia y Tecnolog\'{\i}a of
Spain, under grants FIS2004-05035-C03, FIS2007-65869-C03, and
CTQ2005-00296/PPQ and Comunidad Aut\'onoma de Madrid under program
MOSSNOHO-CM (S-0505/Esp-0299).

\bibliography{biblio}

\end{document}